\documentclass[useAMS,usenatbib]{mn2e}

\usepackage{times}

\usepackage[dvips]{graphicx}
\usepackage{amsmath,amssymb}

\title[Formation of Population III stars in Fossil HII Regions]
{Formation of Population III stars in Fossil HII Regions:\\
Significance of HD}
\author[T. Nagakura and K. Omukai]{Takanori Nagakura$^{1,2}$\thanks{E-mail:
nagakura@th.nao.ac.jp (TN) ; omukai@th.nao.ac.jp (KO)} 
and Kazuyuki Omukai$^{1}$\\
$^{1}$National Astronomical Observatory of Japan, Osawa, Mitaka, 
Tokyo 181-8588 \\
$^{2}$Department of Astronomy,The University of Tokyo,Hongo
7-3-1,Bunkyo-ku,Tokyo 113-0033}
\begin{document}



\maketitle

\label{firstpage}

\begin{abstract}
We study the evolution of gas in HII regions around 
the first stars after the exciting stars end their lives.
If the first star in a small halo dies without supernova (SN), 
subsequent star formation is possible in the same halo.
We thus investigate the effect of ionization
on subsequent star formation within small halos in the early universe 
using one-dimensional hydrodynamics with spherical symmetry 
along with non-equilibrium primordial gas chemistry.
We find that the enhanced electron fraction facilitates the formation of
molecular hydrogen at the cores of these halos.
The low temperature circumstances produced 
by the H$_2$ cooling is favorable to  
HD formation and the resultant cooling 
further drops the temperature below 100 K.
Consequently, low-mass stars with primordial 
abundances can form even in a small halo.
After accreting the interstellar metals,
these stars might resemble low-mass ultra metal-poor stars 
discovered in the present-day Galactic halo.
\end{abstract}


\section{Introduction}
\label{sec:intro}

Observations of high-z quasars indicate that
the intergalactic medium (IGM) was ionized at z $>$ 6.
In addition,
the recent \textit{Wilkinson Microwave Anisotropy
Probe} (WMAP) observations
suggest that the universe was reionized as early as z $\sim$ 17 
(Kogut et al. 2003).
It is still unclear what astrophysical phenomena caused 
the reionization of the universe, 
but photoionization by massive stars is among the most promising candidates.
It is hence important to study the star formation in the early universe.
The first generation of stars was formed out of 
unmagnetized gas of primordial composition, which 
leads to simplification of the relevant physics.
On the other hand, the formation processes of second-generation 
stars are more complex because of various feedback effects 
from the first stars, 
for example, radiative heating and ionization/dissociation, metal enrichment, 
secondary density and velocity fluctuations etc. 
(e.g., Ciardi \& Ferrara 2004).
For reliably predicting the star formation history and the initial mass function
(IMF), etc., it is crucial to understand various feedback effects in the 
high-redshift universe.

In the context of cold dark matter (CDM) cosmology, 
the first collapsed objects are predicted to have masses 
corresponding to virial temperature $T_{\rm vir} < 10^{4}{\rm K}$
(hereafter, we refer to such objects as ``small halos''). 
In such halos, gas cools and contracts further
to form stars only via H$_2$ rovibrational emission. 
The first stars, which formed in small halos, are indicated  
to have masses of a few hundreds M$_{\odot}$
(Abel, Bryan \& Norman 2002; Bromm, Coppi \& Larson 2002;
 Omukai \& Palla 2003).
Once such a massive first star forms in a small halo,
an enormous amount of ionizing photons from the star ionizes its
surrounding gas (Whalen, Abel \& Norman 2004; Kitayama et al. 2004) and 
photodissociates the molecular hydrogen.
Consequently, star formation in the same halo is completely suppressed
(Omukai \& Nishi 1999; Glover \& Brand 2001). 
After that, depending on the mass, the central star 
will explode as a SN or collapse directly into
a BH at the end of its lifetime.
Zero-metal stars in the ranges 10 -- 40 M$_{\odot}$ 
and 140 -- 260 M$_{\odot}$ explode as type II SNe or pair-instability SNe,
respectively (Heger \& Woosley 2002 ; Umeda \& Nomoto 2002).
The ashes of these first SNe pollute their surrounding gas with metals and
dust grains, out of which subsequent generations of low-mass stars form 
(Bromm et al. 2001; Schneider et al. 2002; Omukai et al. 2005).
However, more often than not, the first stars directly form BHs
without SNe   
since the stellar mass range that causes SNe is rather limited.
After the death of the exciting star, ionized gas in the HII region 
begins to recombine. 
We call such an ex-HII region after the death of the exciting star 
a fossil HII region (after Oh \& Haiman 2003).
In shock ionized gas in large halos at virialization and SN shells, 
formation of H$_2$ and HD molecules are promoted 
(Mac Low \& Shull 1986; Shapiro \& Kang 1987; Yamada \& Nishi 1998).
A similar situation can happen also in a fossil HII region.
High ionization degree there facilitates formation of H$_2$ via the H$^-$ 
channel. The gas can cool and contract owing to the H$_2$ cooling, 
thereby producing a subsequent generation of stars in the same halo.
In particular, in those halos, more H$_2$ is formed and then lower temperature 
is achieved than in the first star formation because of the initial ionization.
The enhanced H$_2$ cooling and resultant lower temperature environment 
facilitates HD formation. 
Although the mass of primordial stars is often supposed to be very high,
HD cooling, if it becomes important, enables the formation 
of low-mass stars (Uehara \& Inutsuka 2000; Nakamura \& Umemura 2002).

The problems we try to solve in this paper are the following:
Does the ionized primordial gas in a small halo 
cool and subsequently form stars again inside the same halo?
Does HD work as an efficient coolant in such situations ?
To tackle these problems,
we treat the gas as one-dimensional spherically symmetric fluid and
follow the evolution with solving chemical reactions in primordial gas.

This paper is organized as follows:
Section 2 describes the numerical model used in this paper.
Section 3 presents the results of our simulation,
and Section 4 is devoted to the discussion.
Finally, we summarize our conclusions.
Throughout this paper, we adopt a flat $\Lambda$CDM cosmology
with density parameters $\Omega_{\Lambda}$ = 0.73, 
$\Omega_m$ = 0.27, $\Omega_b$ = 0.044 and the Hubble parameter $h=0.71$
(Spergel et al. 2003).

\section[]{The Model}
\label{sec:models}

As an evolution of fossil HII regions, which were once stellar
ionized regions but now are recombining because of the death of 
the exciting stars, 
within small halos in the early universe, 
we envisage the following scenario. 
First, the first star forms at the center of a small halo.
Then, an enormous amount of ionizing flux from the first star
forms an HII region.
During the lifetime of the exciting star,
the radius of gas accelerated by high pressure of the ionized gas 
reaches 
$\sim c_s t_{{\rm OB}} \sim 
50 ( T/2 \times 10^4 {\rm K} )^{1/2}
   ( t_{\rm OB}/3 {\rm Myr} ) {\rm pc}$, 
where $c_s$, $t_{{\rm OB}}$ are the sound speed of the gas and 
the lifetime of a massive star, respectively, 
while the HII region itself extends to an even larger radius.
Note that despite the shallow potential well of a small halo,  
photoevaporation does not always occur:
since the HII region is smaller than the radius of the halos, 
gas can avoid photoevaporation if cooling takes place immediately.
If the star explodes as a SN at the end of its lifetime,
the gas inside these halos will be swept out by its blast wave
(Bromm, Yoshida \& Hernquist 2003 ; Wada \& Venkatesan 2003).
However, here we consider the case where
the central star directly forms a BH without a SN explosion.
Specifically, we calculate the evolution of the ionized gas 
starting from the state after the death of the central star
(Whalen et al.2004; Kitayama et al.2004).
We employ one-dimensional hydrodynamics 
with spherical symmetry along with primordial gas chemistry.
In this simulation, we ignore any influence of a relic BH, 
which resides in the central region of the halo, on its surrounding gas, 
for simplicity. 
In reality,
accretion onto the BH generates some photodissociative radiation. 
According to O'Shea et al. (2005), however, 
this does not affect the H$_2$ formation in the halo significantly.
The gravitational effect of the BH is assessed from its Bondi accretion 
radius in this paper.  
At the outer boundary,
since our interest is the central region of the gas, 
we take the gas radius to be 200 pc, which is the radius of the expanding
region according to Whalen et al. (2004)'s results, and 
set the pressure outside the cloud to be zero.
This setting means that only the accelerated region inside the HII region is 
considered: the matter outside is neglected whether it is ionized or not.
We have also performed runs with a gas radius of 1 kpc that
corresponds to the 
whole HII region and have confirmed that our results remain almost unaltered 
for the halo mass $\ga$ 10$^6$ M$_{\odot}$.

\subsection{Hydrodynamics}
\label{sec:hydro}

We assume spherical symmetry, 
and neglect both magnetic field and external radiation field except for
cosmic microwave background (CMB) radiation.
The basic equations of one-dimensional spherically symmetric fluid are 
the following:
\begin{align}
&\frac{d M_b}{d r_b} = 4 \pi r_b^2 \rho_b ,
\label{hyd_eq1} \\
&\frac{d^2 r_b}{dt^2} =
- 4 \pi r_b^2 \frac{dp}{dM_b} - \frac{GM(r_b)}{r_b^2}
+ \Omega_{\Lambda} H_0^2 r_b  , \label{hyd_eq2}
\\
&\frac{du}{dt} =\frac{p}{\rho_b^2} \frac{d \rho_b}{dt}
- \frac{\Lambda}{\rho_b} ,  \label{hyd_eq3}
\\
&p=(\gamma -1) \rho_b u=\frac{k_B \rho_b T}{\mu m_p} , \label{hyd_eq4}
\end{align}
where $r_b$, $M_b$, $\rho_b$, $p$, $T$, $u$, $\mu$, $M(r_b)$, 
$G$, $k_B$, $\Omega_{\Lambda}$, $H_0$, $\Lambda$  
are the radius, mass, 
density, pressure, temperature, internal energy per unit mass, 
mean molecular weight in units of the proton mass $m_p$, 
the total mass interior to $r_b$, the gravitational constant, 
the Boltzmann constant, the cosmological constant, 
the Hubble parameter at the present-day,
and the  cooling rate per unit volume, 
respectively.
The last term of the right-hand side in the second equation represents 
the correction of the expanding universe.
The adiabatic index $\gamma$, which depends on chemical abundances, 
is calculated at each time-step.
We here neglect the chemical heating/cooling term
because it is not important 
in the density range we considered ($<10^8 {\rm cm}^{-3}$).

We solve these equations using the standard second-order-accurate Lagrangian 
finite-difference scheme with von Neumann-Richtmyer artificial viscosity
(e.g., Mihalas \& Mihalas 1984; Thoul \& Weinberg 1995; Omukai \& Nishi 1998).
The grids are spaced unequally in mass. 
The innermost cell contains 3.7 $\times$ 10$^{-6}$ M$_{\odot}$, and 
the increments in radius between adjacent grid cells are increased 
by 5 \% outwardly.
We take 100 cells and have confirmed that it is enough to follow 
the evolution near the center.

\subsection{Dark Matter}
\label{sec:DM}

We need to determine the dark-matter distribution
to solve a set of hydrodynamic equations.
We assume that the dark-matter distribution is fixed in time 
since dark matter has already virialized at the phase that 
we are interested in. 
We adopt the dark-matter density profile derived by 
Navarro, Frenk \& White (1997):
\begin{equation}
\rho(r)=\frac{\rho_{\rm crit} \delta_c}{r/r_s(1+r/r_s)^2} ,
\end{equation}
where $\rho_{\rm crit}$, $\delta_c$, $r_s$ are the critical density at
virialization redshift, characteristic dimensionless density at the
center and characteristic radius, respectively.
This density profile has the following three parameters:
the mass of a dark matter halo M$_{\rm halo}$,
the virialization redshift $z_{\rm vir}$ of the halo, 
and the concentration parameter $c$, which expresses the concentration 
of mass near the center of the halo.
In particular, by using $c$, the virial radius of the halo 
$r_{\rm vir}=c r_{s}$.
Here we fix $c$ = 5 because its small variation does not affect our results.

\subsection{Cooling Processes}
\label{sec:cool_func}

We incorporate rovibrational line cooling of H$_2$ and HD, Ly$\alpha$
 line cooling of atomic hydrogen,
and Compton cooling by CMB radiation   
since we consider the temperature range $<$ 2 $\times$ 10$^4$ K.
For line cooling of H$_2$ we adopt the fitting function given by 
Galli \& Palla (1998).
HD cooling rate is calculated by solving the statistical equilibrium between 
the first 5 rotational levels.
The radiative and collisional transition rate coefficients are
the same as those in Galli and Palla (1998).
Ly$\alpha$ line cooling rate of atomic hydrogen is given by 
(Dalgarno \& McCray 1972)
 \begin{align}
\Lambda_{\rm Ly\alpha} = 7.5 \times 10^{-19} 
\exp \left( - \frac{1.18 \times 10^{5}{\rm K}}{T} \right) n(e) n({\rm
  H}) & \notag \\
{\rm erg~s^{-1}~cm^{-3}} &,
\end{align}
where $n(H)$ is the number density of neutral hydrogen.
The Compton cooling rate is given by 
(e.g.,Rybicki \& Lightman 1979)
\begin{align}
\Lambda_{\rm Compton} &= 
\frac{4 k_B c \sigma_T n(e) U_{\gamma} (T-T_{\gamma})}{m_e c^2} \notag \\
&= 1.017 \times 10^{-37} T_{\gamma}^{4} (T - T_{\gamma}) n(e) \quad {\rm erg~s^{-1}~cm^{-3}},
\end{align}
where $c$, $\sigma_T$, $n(e)$, $U_{\gamma}$, $T_{\gamma}$ are 
the speed of light, Thomson scattering cross section, 
number density of electron, 
energy density of photon and photon temperature, respectively.
When the gas temperature decreases as low as the CMB temperature 
$T_{\gamma}$, the heating by the radiation becomes important.
We include this effect by replacing the radiative cooling rate $\Lambda$
with $\Lambda (T) - \Lambda (T_{\gamma})$. 

\subsection{Chemical Reactions}
\label{sec:chem}

We need to solve chemical reactions as well as 
hydrodynamic equations
since cooling of the gas depends on chemical abundances of the coolants.
We assume that the reactions are only two-body collisional ones and
do not include the three-body H$_2$ formation process
because we consider only the low density range ($n_H$ $<$ 10$^8$ cm$^{-3}$).
The evolution of chemical species $i$ is followed by solving the kinetic
equation
\begin{equation}
\frac{dn(i)}{dt}
= - n(i) \sum_{j} k_{ij} n(j) + \sum_{l,m} k_{lm} n(l) n(m),
\label{chem_eq1}
\end{equation}
where $k_{ij}$ and $k_{lm}$ are the reaction rate coefficients 
of destruction and those of formation, respectively.

Since no pollution of the gas with metals has taken place in this situation,
we consider the following 14 species:
 H, H$^{+}$, H$^{-}$, H$_2$, H$_2^{+}$, He, He$^{+}$, 
He$^{++}$, e$^{-}$, D, D$^{+}$, D$^{-}$, HD, HD$^{+}$.
The fractions of He and D atoms are assumed to be 0.0833 and  
2.5 $\times$ 10$^{-5}$, respectively, which are inferred from 
big bang nucleosynthesis (Romano et al. 2003). 
Reactions among H and He compounds and their rate coefficients 
are the same as those in Abel et al. (1997).
For D reactions, we adopt the compilation by Nakamura \& Umemura (2002), 
although we replace the rate coefficients of the reactions D5, D9, D10 and D11 
with those by Galli \& Palla (2002).
The total number of considered reactions amounts to 37.
We solve the rate equations for 14 species above 
using an implicit difference scheme.

\subsection{Initial and Boundary Conditions}
\label{sec:condition}

We assume that the first star in a small halo is formed around the time
of virialization.
The initial epoch of our calculation is after 3 Myr, 
the typical lifetime of massive stars,  of virialization.
The evolution of the first HII regions in the early universe has been
studied by Whalen et al. (2004) and Kitayama et al. (2004).
We construct the initial density and velocity distributions following
Whalen et al. (2004)'s results.
Their results show that, inside the HII region, 
there are two distinct regions:
the inner region is expanding with outwardly increasing velocity and
the velocity reaches the sound speed 
of the ionized gas at a radius of $\sim$ 200 pc, while   
the gas outside is almost unaccelerated. 
We consider only the expanding region in our calculation.
Also, according to Whalen et al. (2004), 
the density distribution inside that region is almost constant 
($\sim$ 0.06 cm$^{-3}$), 
and the velocity increases gradually from the center outward.
As the fiducial runs, we assume that the density profile of 
gas is constant at 0.06 cm$^{-3}$,
and that the velocity is proportional to the radius.
At 200 pc from the center, the velocity is about sound speed. 
There is no gas outside this region.
We set the initial gas temperature inside the HII region to be 2
$\times$ 10$^4$ K.  
The gas is assumed to be fully ionized initially.

This initial condition corresponds to the total baryonic mass of 
$6 \times 10^{4} M_{\odot}$ in the fiducial cases.
In the case of 10$^5$ M$_{\odot}$ halos,
this condition leads to too much baryon compared to dark matter.
Probably a lower gas-density state is more appropriate as an initial 
condition for such a small halo.
Our choice of the initial condition facilitates the cloud 
to cool and contract. 
Even with this favorable condition, however, they do not cool anyway 
as described later.
Then, the results do not change even if the initial gas density is lowered.
For simplicity, we use the same initial condition for baryons 
in all the cases.

The size of the flat density region ($\sim$ 200 pc) 
is basically the sound speed in the HII region $c_{\rm s}$
times the lifetime of the exciting stars $t_{\rm OB}$.
Although the size of the flat density region does not depend 
significantly on halo quantities, 
there should still be some density variations among halos.
To study the effect of different initial gas densities, 
we also calculate the cases where the initial gas density is 
either halved or doubled from the fiducial value.

We follow the evolution until the local Hubble time at $z_{\rm vir}$ 
passes or the central number density
reaches the threshold value 10$^8$ cm$^{-3}$.
When the latter condition is satisfied, we regard the gas to have 
cooled and collapsed.
We take the threshold density at 10$^8$ cm$^{-3}$, since
three-body H$_2$ formation, which we do not include here, 
becomes important for higher densities 
and the evolution cannot be followed correctly thereafter.
The results do not depend on the value of the threshold density as 
long as the free-fall time there is sufficiently shorter than 
the local Hubble time.   

\section{Results}
\label{sec:results}

We simulate the evolution of ionized primordial gas in small halos 
of different masses and formation redshifts.
The parameters of the runs that we have performed are listed 
in Table \ref{tab:vir}.

\begin{table}
 \begin{center}
  \begin{tabular}{cccccc}
\hline 
   $z_{{\rm vir}}$ & M$_{{\rm halo}}$ [M$_{\odot}$] 
& T$_{{\rm vir}}$ [K] & $\Delta$ t [yr] & z$_{\rm end}$ & cooled ? \\
\hline
   15
 & 10$^5$ & 1.6 $\times$ 10$^2$ & -- & -- & No \\
 & 10$^6$ & 7.6 $\times$ 10$^2$ & 1.65 $\times$ 10$^{8}$ & 10.7 & Yes \\
 & 10$^7$ & 3.5 $\times$ 10$^3$ & 1.35 $\times$ 10$^{7}$ & 14.4 & Yes \\
 & 10$^8$ & 1.6 $\times$ 10$^4$ & 4.67 $\times$ 10$^{6}$ & 14.7 & Yes \\
\hline
   20 
 & 10$^5$ & 2.1 $\times$ 10$^2$ & -- & -- & No \\
 & 10$^6$ & 1.0 $\times$ 10$^3$ & 5.25 $\times$ 10$^{7}$ & 16.6 & Yes \\
 & 10$^7$ & 4.6 $\times$ 10$^3$ & 7.76 $\times$ 10$^{6}$ & 19.2 & Yes \\
 & 10$^8$ & 2.1 $\times$ 10$^4$ & 3.08 $\times$ 10$^{6}$ & 19.6 & Yes \\
\hline
   25 
 & 10$^5$ & 2.7 $\times$ 10$^2$ & -- & -- & No \\
 & 10$^6$ & 1.2 $\times$ 10$^3$ & 2.56 $\times$ 10$^{7}$ & 21.9 & Yes \\
 & 10$^7$ & 5.7 $\times$ 10$^3$ & 5.44 $\times$ 10$^{6}$ & 23.9 & Yes \\
 & 10$^8$ & 2.7 $\times$ 10$^4$ & 2.29 $\times$ 10$^{6}$ & 24.3 & Yes \\
\hline
   30 
 & 10$^5$ & 3.2 $\times$ 10$^2$ & -- & -- & No \\
 & 10$^6$ & 1.5 $\times$ 10$^3$ & 1.59 $\times$ 10$^{7}$ & 26.7 & Yes \\
 & 10$^7$ & 6.8 $\times$ 10$^3$ & 4.23 $\times$ 10$^{6}$ & 28.6 & Yes \\
 & 10$^8$ & 3.2 $\times$ 10$^4$ & 1.82 $\times$ 10$^{6}$ & 29.1 & Yes \\
\hline
  \end{tabular}
  \caption{The list of our runs. We list the
  virialization redshift, the halo mass, the virial temperature,
  the time elapsed in our simulation, the redshift at the end of 
  our simulation,
  and whether the gas can cool or not.}
  \label{tab:vir}
 \end{center}
\end{table}%

\begin{figure*}
\begin{center}
 \scalebox{1.0}{
  \includegraphics[width=\hsize]{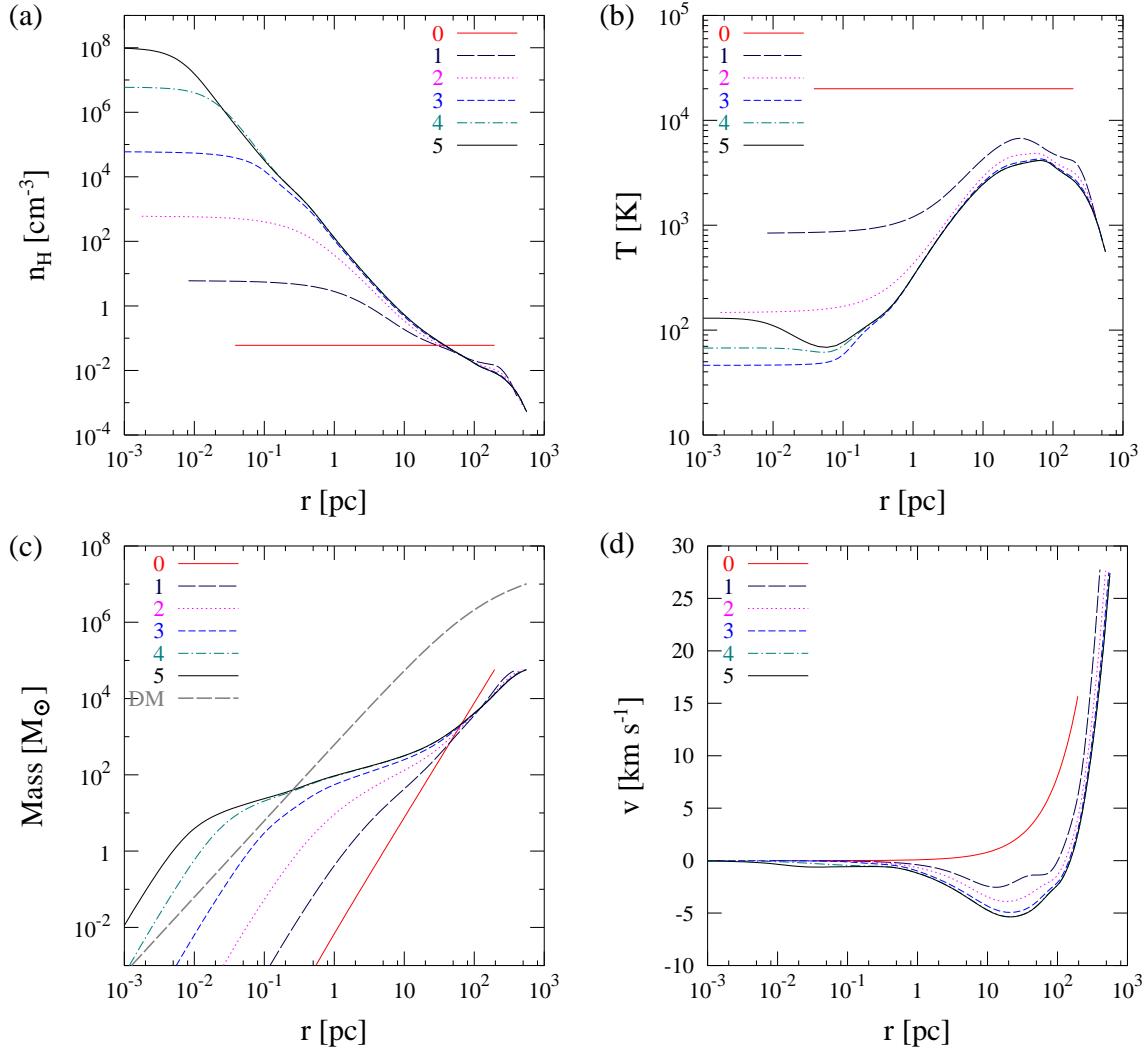}
}
\end{center}
 \caption{The evolution of the number density (a), temperature (b), mass
 (c),
 velocity distribution (d) as a function of the radius for the fiducial run
(z$_{{\rm vir}}$=15, M$_{{\rm halo}}$=10$^{7}$ M$_{\odot}$).
 The numbers represent the evolutionary sequence, and we set 
 our initial state as state 0. The numbers
 from 1 to 5 correspond to the states after 7.94 Myr, 2.90 Myr, 1.93 Myr, 713
 kyr, 47.7 kyr from the previous ones.
 Each state we numbered has central density 100 times larger than that of 
 the previous state.
 In panel (c), we also plot the dark-matter mass distribution 
(indicated as ``DM'').
 The initial low gas density $(\sim 0.1 {\rm cm^{-3}})$ reflects 
 our choice of the initial state, which is the gas inside 
 the HII region. 
 Since we consider only the matter inside the radius of 200 pc, 
 the total gas mass in our calculated region is lower than 
 that in the halo 
 $(\Omega_b/\Omega_m)M_{\rm halo} \sim 1.7 \times 10^6 M_{\odot}$.
 }
 \label{fig:157rth}
\end{figure*}%

\begin{figure*}
\begin{center}
\scalebox{1.0}{
  \includegraphics[width=\hsize]{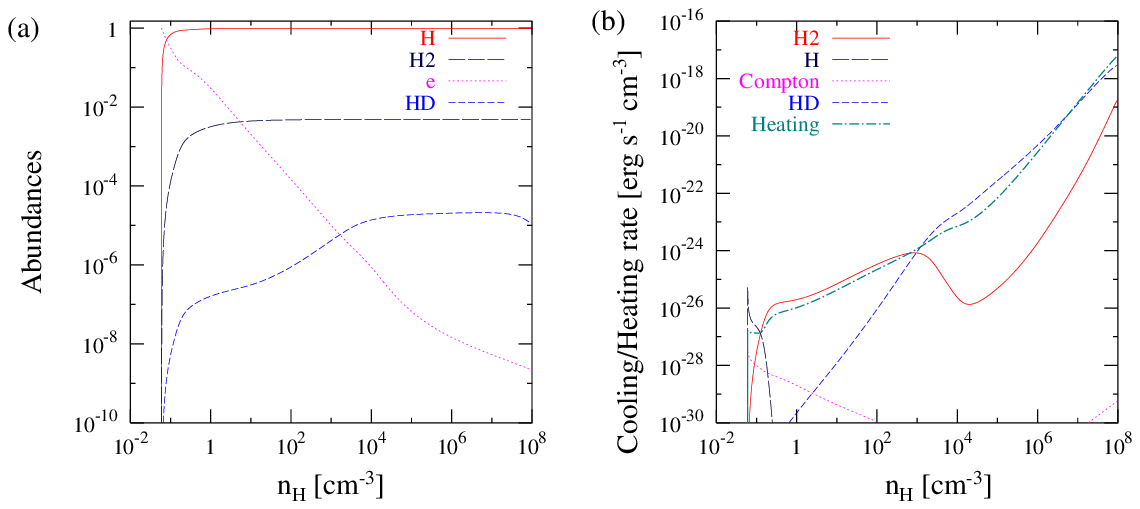}
}
\end{center}
 \caption{The evolution of the chemical species (a) and the
 cooling/heating rates (b) at the center for the fiducial run 
(z$_{{\rm vir}}$=15,M$_{{\rm halo}}$=10$^{7}$ M$_{\odot}$). 
In panel (a), the H, H$_2$, HD and electron fractions at the center
is shown as a function of the central number density.
In panel (b), the H Ly$\alpha$ cooling, the H$_2$ cooling rate, 
the HD cooling rate, the Compton cooling and the compressional 
heating rate per unit volume are plotted.}
 \label{fig:157nth}
\end{figure*}

\begin{figure}
 \begin{center}
  \includegraphics[width=10cm]{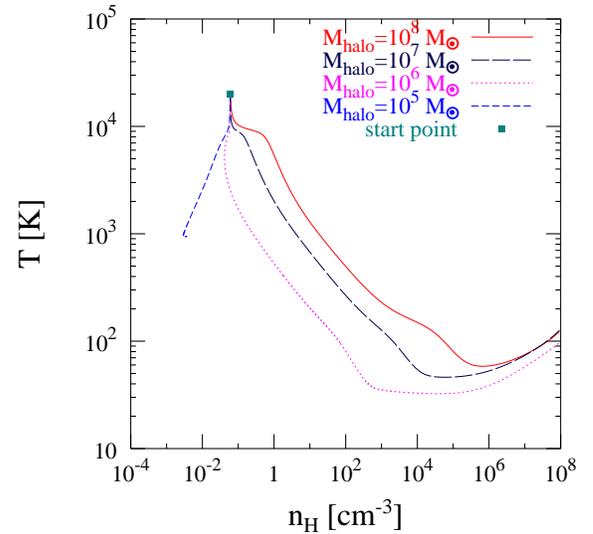}
  \end{center}
 \caption{The evolution of the central temperature as a function of 
 the central number density for halos with z$_{{\rm vir}}$ = 15. 
 We plot the cases of 
 M$_{{\rm halo}}$ = 10$^5$, $10^6$, 10$^7$, and 10$^8$ M$_{\odot}$.
 The square represents the initial state.
}
 \label{fig:15nt}
\end{figure}

We first show the case where the dark halo virializes at
z$_{\rm vir}$ = 15 and its mass M$_{\rm halo}=10^7$
M$_{\odot}$. Hereafter we call this case our fiducial model. 
In Figure \ref{fig:157rth}, we show the evolution of
the number density, temperature, mass, and velocity distributions.
The numbers from 0 to 5 in this figure represent the time sequence.
State 0 corresponds to the initial distribution. 
The numbers are assigned every time when the central density reaches
100 times that of the previous numbered state.
The numbers from 1 to 5 correspond to the states after  
7.94 Myr, 2.90 Myr, 1.93 Myr, 713 kyr and 47.7 kyr 
of the previous state, respectively.
The time passed from 0 to 5 is 13.5 Myr and $z$ $\approx$ 14.4 at state
5.
The central evolution is shown in Figure \ref{fig:157nth}.
In this figure, we plot the evolution of chemical abundances
and cooling/heating rates as functions of the central number
density.
In Figure \ref{fig:15nt},
we plot the central temperature against the
central number density along with cases of different halo sizes.
In the fiducial case, the number density at the center reaches 
10$^8$ cm$^{-3}$
within the local Hubble time at $z$ = 15 ($\sim 4.1 \times 10^8$ yr),
namely, the cloud has cooled and collapsed successfully.

Without dark matter, the density distribution of 
collapsing clouds is known to follow the Larson-Penston-type similarity 
solution (Larson 1969; Penston 1969). 
This solution consists of two regions, i.e., a core with flat density 
distribution and an envelope with outwardly decreasing density.
The diameter of the core is approximately the Jeans length 
$\lambda_J$ $\sim$ $c_s / \sqrt{\rho_{\rm gas}}$.
In our case, we need to consider the dark matter gravity.
In Figure \ref{fig:157rth} (a), we see that the density distribution is
similar to the Larson-Penston type solution, but the core radius is smaller.
Owing to the dark matter contribution, 
the core radius is about $c_s / \sqrt{\rho_{\rm gas} + \bar{\rho}_{\rm DM}}$, 
where $\bar{\rho}_{\rm DM}$ is the dark-matter density averaged in the core.
This means that dark matter reduces gas density in the collapsing 
clouds.
The small bent in the density distribution at $\sim 0.1$ pc corresponds
to the transition from the dark matter to the self-gravity dominance. 

A large amount of molecular hydrogen is immediately formed  
(Figure \ref{fig:157nth} a) through the H$^{-}$ channel
\begin{eqnarray}
&{\rm H} + {\rm e}^{-} \rightarrow {\rm H}^{-} + \gamma ,
\label{eq:H2_1} \\
&{\rm H} + {\rm H}^{-} \rightarrow {\rm H}_2 + {\rm e}^{-} ,
\label{eq:H2_2}
\end{eqnarray}
because the electron, which is the catalyst of this reaction, 
is abundant in our initial state.
On the other hand, the electron fraction declines dramatically through
recombination with increasing density.
Because of the reduced number of electrons, 
the abundance of molecular hydrogen becomes saturated. 
Around the center, the H$_2$ fraction finally reaches 
about 5 $\times$ 10$^{-3}$, which 
is well above that in the case of a halo without initial
ionization (about 10$^{-4}$ for  $n_{\rm H} \la$ 10$^{4}$ cm$^{-3}$; 
e.g., Abel et al. 2002).
In our case, the temperature drops enough to trigger HD formation.
The resultant HD cooling reduces the temperature below 100 K (see Figures 
 \ref{fig:157rth} b and \ref{fig:15nt}).
The total mass of the region that is eventually cooled below 100 K 
amounts to $\sim$ 28 M$_{\odot}$.

Note that the gas in small halos with T$_{{\rm vir}}$ $<$ 10$^4$ K will 
be photoevaporated unless it cools below T$_{\rm vir}$ by molecular 
cooling within a sound-crossing time.
Our results indicate that, even in such small halos, some clouds 
can contract without photoevaporation owing to 
the instant effect of H$_2$ cooling.

Figure \ref{fig:157nth} (b) shows the contribution to the cooling rate by 
important coolants in the course of the contraction.
Also shown is the compressional heating rate. 
At the initial state, because the gas temperature is 2 $\times$ 10$^4$ K, 
the Ly$\alpha$ line cooling of atomic hydrogen is the most effective.
Below 10$^4$ K the Ly$\alpha$ line cooling is cut off since the atomic
hydrogen becomes very hard to excite.
Then, H$_2$ cooling, which compensates compressional heating subsequently,
enables further contraction of the gas.
When the temperature decreases below 200 K at 10$^3$ cm$^{-3}$, 
molecular hydrogen is no longer excited and its cooling rate declines rapidly.
The low temperature at this point is sufficient to trigger abundant HD
production (see Figure \ref{fig:157nth} a ).

The HD abundance is determined by equilibrium between the formation rate
and its reverse rate (e.g., Omukai et al. 2005),
\begin{equation}
D7, D9: {\rm D}^{+} + {\rm H}_{2} \rightleftarrows {\rm H}^{+} + {\rm HD}.
\end{equation}
By equating rates of these reactions,  
\begin{equation}
\frac{n({\rm HD})}{n({\rm H}_2)} 
= \frac{k_{D7}n({\rm D}^{+})}{k_{D9}n({\rm H}^{+})}, 
\label{eq:HD79}
\end{equation}
where $k_{D7}$ and $k_{D9}$ are the forward and its reverse reaction
rate coefficients respectively.
The abundance of D$^{+}$ is given by the equilibrium between the 
following reactions:
\begin{equation}
D2, D3: {\rm D} + {\rm H}^{+} \rightleftarrows {\rm D}^{+} + {\rm H}.
\end{equation}
Similarly, 
\begin{equation}
\frac{n({\rm D^{+}})}{n({\rm H^{+}})}
=\frac{k_{D2} n({\rm D})}{k_{D3} n({\rm H})}.
\label{eq:HD23}	
\end{equation}
From equations (\ref{eq:HD79}) and (\ref{eq:HD23}), 
the HD to H$_{2}$ ratio is given by 
\begin{equation}
\frac{n({\rm HD})}{n({\rm H}_2)} 
= \frac{k_{D2} k_{D7} n({\rm D})}{k_{D3} k_{D9} n({\rm H})}
= 2 \exp \left( \frac{421 {\rm K}}{T} \right) \frac{n({\rm D})}{n({\rm H})}.
\label{eq:HDabundance}
\end{equation}
For the numerical value, we used the rate coefficients compiled by  
Nakamura \& Umemura (2002). 
The equation (\ref{eq:HDabundance}) tells that 
because the binding energy of HD is higher than that of H$_{2}$ 
by $\Delta E/k_{\rm B}=421$K, HD is a state preferable to 
H$_2$ at the low temperature.
Therefore the HD formation is promoted in an environment 
where the temperature is much lower than the difference of binding energies. 

After the temperature falls below 200 K,
HD becomes the dominant coolant and its cooling exceeds the 
compressional heating.
Then the temperature continues to decrease and eventually 
reaches below 100 K.
This trend continues until the number density 10$^5$
cm$^{-3}$, 
which is the critical density for HD to reach the local thermodynamic
equilibrium (LTE).
The HD cooling rate per unit mass subsequently saturates and 
temperature begins to increase with  
the compressional heating now exceeding the HD cooling (see Figure 
\ref{fig:157nth} b).

Let us evaluate mass of the cooled region at the center of the cloud, 
which is expected to become stars eventually.
For gas to contract and fragment into stars,
it must be bounded by its self-gravity.
The condition for the gas self-gravity dominating the dark matter
gravity within the radius $r$ is 
that the enclosed gas mass $M_{{\rm gas}}(r)$ is larger than the enclosed
dark matter mass $M_{{\rm DM}}(r)$.
We hereafter refer to a central portion satisfying the condition above 
as a central cooled region.
For the fiducial run, we plot the gas mass and the dark matter mass 
as a function 
of the radius in Figure \ref{fig:157nth} (c).
From this, we see that the radius of the central cooled region is 
about 0.27 pc and the gas mass contained there is about 45 M$_{\odot}$.

\begin{figure*}
 \begin{center}
  \scalebox{1.0}{
  \includegraphics[width=\hsize]{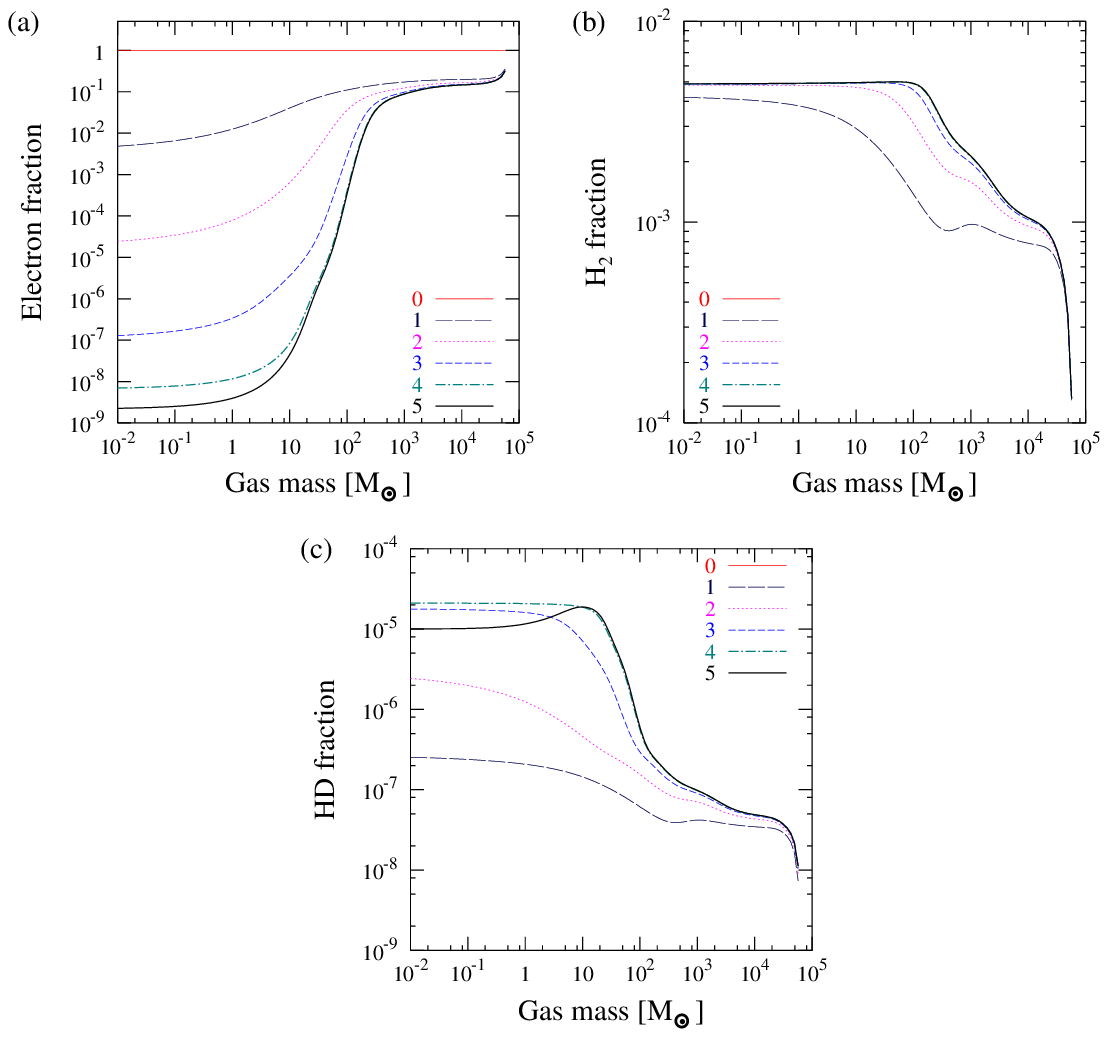}
  }
 \end{center}
 \caption{The evolution of the electron, H$_2$, and HD fraction 
as a function of the Lagrangian gas-mass coordinate for the fuducial 
run (z$_{{\rm vir}}$=15 and M$_{{\rm halo}}$=10$^{7}$ M$_{\odot}$).
The numbers have the same meaning as those in Figure \ref{fig:157rth}.}
 \label{fig:157mch}
\end{figure*}

In Figure \ref{fig:157mch},
we show the evolution of H$_2$, HD and electron
distributions.
The region of high H$_2$ fraction ($\ga 10^{-3}$) is as large 
as $10^{4}M_{\odot}$. 
Within the central $\sim 100M_{\odot}$, $y({\rm H_2})$ reaches 
$5\times 10^{-3}$, while the electron is significantly depleted.
In the central $\sim 10M_{\odot}$, D becomes fully molecular. 
In the outermost region, although the electron is abundant, 
H$_2$ formation is suppressed because of very low density.
The temperature decrease ($< 10^4$ K) there is caused by the expansion of
the gas as well as by the Ly$\alpha$ line cooling and the Compton cooling. 

We also calculated the cases where the initial gas density is halved
and doubled.
In such cases, the gas evolution is qualitatively very similar to our 
fiducial case. 
The size and mass of the central cooled regions in 
these cases are also similar to those of the fiducial case.
Their differences from the fiducial case are within a factor of 2 in mass.
In the halved density case
the size and mass are $\sim$ 0.22 pc and $\sim$ 31 $M_{\odot}$, 
while in the doubled density case  
they are $\sim$ 0.35 pc and $\sim$ 77 $M_{\odot}$, respectively.
We conclude that the effect of different initial density 
on the size and mass of the cooled region is small, 
although present to some extent.

Next, we discuss the run with $z_{{\rm vir}}$ = 15 and
M$_{{\rm halo}}$ = 10$^8$ M$_{\odot}$, 
where the dark halo mass is ten times larger than our fiducial one.
The central temperature evolution has been presented in Figure 
\ref{fig:15nt}.
Also in this case, the central number density reaches 10$^8$ cm$^{-3}$ 
within the Hubble time at $z_{{\rm vir}}$ = 15:
by our criterion, the cloud has cooled and collapsed.
The density and temperature evolve in a fashion similar to that of the
fiducial case.
The time required to collapse is, however, several times smaller 
because of the deeper potential well:
the total elapsed time is about 4.7 Myr and $z$ $\sim$ 14.7 at the final state.
The temperature, chemical abundances, and cooling/heating rates 
also qualitatively behave as in the fiducial run, but
the cooling/heating rates are larger by two orders of magnitude.
This is because the shorter collapse time and the consequent higher
compressional heating keep the temperature higher than in the fiducial
run (see Figure \ref{fig:15nt}).
By $n_{\rm H}$ $\ga$ 10$^6$ cm$^{-3}$ the evolution of temperature and then
cooling rate has converged with that in the fiducial run.  
The evolution in the cases of smaller halo masses with the same virialization
redshift is as follows: 
For a halo with 10$^6$ M$_{\odot}$, the gas cools and collapses 
within the local Hubble time, 
while that in a halo with 10$^5$ M$_{\odot}$ does not.
For the halo of 10$^5$ M$_{\odot}$ and initially also for that of 10$^6$
M$_{\odot}$, 
the number density decreases because of the initial expanding motion.
For the halo with 10$^6$ M$_{\odot}$, 
the initially expanding gas eventually cools and contracts enough 
within the Hubble time
while for the halo with 10$^5$ M$_{\odot}$ the gas continues expanding 
owing to shallow potential well.

\begin{figure}
 \begin{center}
  \includegraphics[width=10cm]{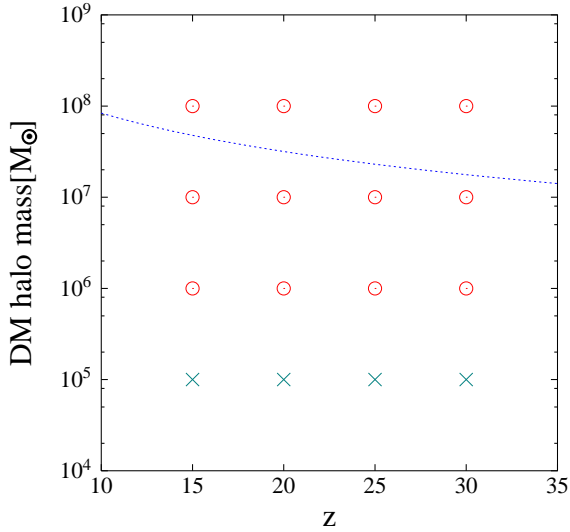}
 \end{center}
 \caption{The $z_{\rm vir}$ -- $M_{\rm halo}$ diagram showing 
 whether the ionized gas in 
 the halos of virialization redshift $z_{\rm vir}$ and mass $M_{\rm halo}$
 can cool and collapse or not.
 The cases where the gas cools and collapses within the local Hubble time 
 are shown by the circles, while the other cases are by the crosses. 
 The dotted line represents T$_{{\rm vir}}$ = 10$^4$ K,  
 objects below which are ``small halos''.
 }
 \label{fig:coll_Mz}
\end{figure}

So far, we have seen the results for halos
virializing at z = 15.
The case for the halo virializing earlier ($z = 30$) with 
smaller mass (10$^6$ M$_{\odot}$) closely resembles our fiducial run.
This is because the earlier a halo virializes, the more 
dark matter tends to concentrate at the center of a halo.
Consequently, 
increasing $z_{{\rm vir}}$  cancels decreasing M$_{{\rm halo}}$. 
We perform additional runs with z$_{{\rm vir}}$ = 20 and 25
for M$_{\rm halo}$ = 10$^5$, 10$^6$, 10$^7$ and 10$^8$ M$_{\odot}$ 
in order to see whether or not the ionized gas 
within these halos cools and collapses.
We summarize the results for all runs in Figure \ref{fig:coll_Mz} 
and in Table \ref{tab:vir}.
Roughly speaking, the gas in a fossil HII region 
collapses and forms stars again 
if the halo mass is larger than 10$^6$ M$_{\odot}$.
We also summarize the size and the gas mass of the central cooled
regions for all runs in Table \ref{tab_cool}.
The size of the central cooled region becomes smaller with increasing 
the halo mass because the collapse time-scale becomes shorter 
(Table \ref{tab:vir}) and there is not enough time to cool.

Here we have included only the initially expanding gas in our simulation
boundary. 
In reality, a large amount of the static material exists just outside the 
boundary.
If we include it, the expanding gas will collide with it and 
will be decelerated.
Then, the collapse of the gas is facilitated.
Consequently, even the gas in halos smaller than $10^{6}M_{\odot}$, 
which could not cool in our runs, might be able to collapse and form stars 
in more realistic calculations.

\begin{table}
 \begin{center}
  \begin{tabular}{cccc}
\hline 
   $z_{{\rm vir}}$ & M$_{{\rm halo}}$ [M$_{\odot}$] 
& Radius [pc] &  Mass [M$_{\odot}$] \\
\hline
   15
 & 10$^6$ & 0.31 & 30 \\
 & 10$^7$ & 0.27 & 45 \\
 & 10$^8$ & 0.16 & 35 \\
\hline
   20 
 & 10$^6$ & 0.28 & 40 \\
 & 10$^7$ & 0.19 & 39 \\
 & 10$^8$ & 0.11 & 28 \\
\hline
   25 
 & 10$^6$ & 0.24 & 46 \\
 & 10$^7$ & 0.14 & 35 \\
 & 10$^8$ & 0.09 & 27 \\
\hline
   30 
 & 10$^6$ & 0.22 & 50 \\
 & 10$^7$ & 0.12 & 34 \\
 & 10$^8$ & 0.07 & 27 \\
\hline
  \end{tabular}
  \caption{The radius and mass of the central cooled region in our runs. 
  From the left, we list the virialization redshift, the halo mass, 
 the radius and the gas mass of the central cooled region.}
  \label{tab_cool}
 \end{center}
\end{table}%

\section{Discussions}
\label{sec:discuss}

We have shown that the temperature in a fossil HII region of the first star
drops as low as the CMB temperature by HD cooling 
even with the primordial composition.
In the case of the formation of first-generation stars, HD does not 
become important and 
such a low temperature is not attained (Bromm et al. 2002).
In our situation, the high electron fraction enables molecular
hydrogen to form in a larger amount than in an initially un-ionized cloud.
Abundant H$_2$ molecules thus formed cool the gas below the threshold 
temperature for HD formation ($\sim 150$K; Omukai et al. 2005).
Under such a low temperature environment, with further contraction
the gas is expected to fragment into low-mass 
($\sim$ 0.1 -- 2 M$_{\odot}$) gas clumps
(Uehara \& Inutsuka 2000; Nakamura \& Umemura 2002).    
Uehara \& Inutsuka (2000) claimed that subsolar mass stars are formed by HD
cooling in a cooled layer behind an ionizing shock due to either 
virialization or a SN.
In the case of a virialization shock, the halo must be massive enough 
to cause a strong shock.
When such large halos are formed in the standard bottom-up scenario of
structure formation,
star formation and metal-enrichment may have proceeded in building blocks.
Similarly, in the case of SN shocked layers, metal-pollution might have
taken place at the explosion.
Therefore, it is not certain whether the gas in such shock-ionized region 
remains metal-free.
On the other hand, here we have proposed a viable scenario for 
forming low-mass stars from primordial pristine gas.
If those low-mass stars accrete metal-polluted interstellar medium after
the formation, they might resemble ultra metal-deficient stars  
(e.g., Yoshii 1981).
This may explain the origin of the ultra metal-deficient low-mass
stars discovered
recently (Christlieb et al. 2002; Frebel et al. 2005).

Here, we estimate the number of second-generation stars forming in a 
fossil HII region.
The mass of the central cooled clump 
M$_{\rm clump}$ is about 45 M$_{\odot}$ in the fiducial run (Section 3).
By using the star formation efficiency $\epsilon_{\ast}$ and 
the typical mass-scale of the second-generation stars M$_{\ast}$,
the number of the second-generation stars is 
\begin{equation}
N_{\ast} = \epsilon_{\ast} M_{\rm clump} / M_{\ast} 
= 5 \left(\frac{\epsilon_{\ast}}{0.1} \right) 
\left(\frac{M_{\rm clump}}{50M_{\odot}}\right)
\left(\frac{M_{\ast}}{1M_{\odot}}\right)^{-1}.
\end{equation}
A few second-generation stars are formed per each first
star if they are in relatively massive ($\ga 10^{6} M_{\odot}$) 
halos.

\begin{figure}
 \begin{center}
  \includegraphics[width=10cm]{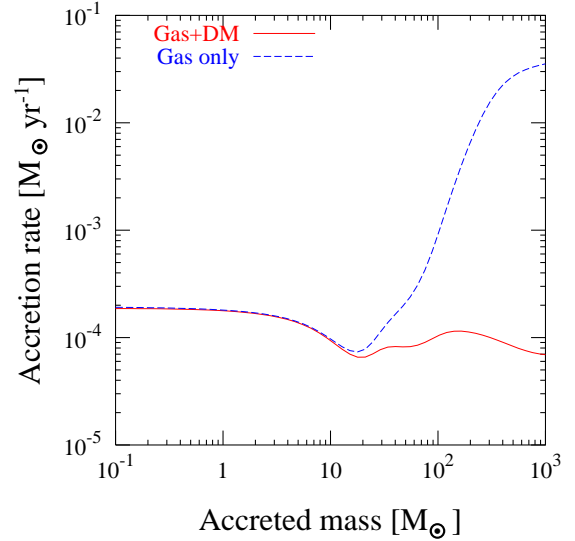}
 \end{center}
\vspace*{0.1cm}
 \caption{Evolution of the mass accretion rate onto the protostar 
forming in the fossil HII region (solid: equation \ref{eq:mdot}).
The instantaneous mass of the protostar is used as a time coordinate.
Also shown is that without dark-matter gravity (dashed).
 }
 \label{fig:mdot}
\end{figure}

Next, we evaluate the accretion rate onto the protostars.
As mentioned in Section \ref{sec:results},
the core radius becomes $r_{\rm core} 
\sim c_s / \sqrt{\rho_{{\rm gas}} + \bar{\rho}_{\rm DM}}$ 
in the presence of dark matter.
Assuming the gas mass in the core 
$M_{\rm gas}\sim \rho_{{\rm gas}} r_{{\rm core}}^3$ 
falls into the center in the free-fall time
$t_{{\rm ff}} \sim 1/\sqrt{G (\rho_{{\rm gas}}+\bar{\rho}_{\rm DM})}$,
the accretion rate becomes
\begin{equation}
\label{eq:mdot}
\dot{M} \sim \frac{M_{{\rm gas}}}{t_{{\rm ff}}} 
\sim \frac{\rho_{{\rm gas}} r_{{\rm core}}^3}{\{G (\rho_{{\rm gas}}+
\bar{\rho}_{\rm DM})  \}^{-1/2}} 
\sim \left( \frac{\rho_{{\rm gas}}}{\rho_{{\rm gas}}+\bar{\rho}_{\rm DM}} 
\right) \frac{c_{\rm s}^3}{G}.
\end{equation}
Namely, when dark matter dominates the gravity, 
the accretion rate becomes smaller than that without dark matter
by a factor of $\rho_{{\rm gas}} /(\rho_{{\rm gas}}+ \bar{\rho}_{\rm DM})$. 
In Figure \ref{fig:mdot}, we show the mass accretion rate by 
equation (\ref{eq:mdot}) for the fiducial run. 
As a time coordinate, we use the mass of the central star, i.e., 
the gas mass that has already accreted at a given time.
Also shown for comparison is $c_{\rm s}^3/G$, which is the accretion rate
without dark matter (Shu 1977).
Initially, the effects of dark matter is weak in the accretion rate
because the self-gravity dominates in the inner part. 
The accretion rate is about $10^{-4}M_{\odot}/{\rm yr}$.
This is smaller than that in the first star formation,  
$\sim 10^{-3}-10^{-2}M_{\odot}/{\rm yr}$ because of 
the lower temperature resulting from HD cooling.
Later the discrepancy from $c_{\rm s}^3/G$ becomes significant 
because of dominant dark-matter gravity:
the accretion rate by equation (\ref{eq:mdot}) remains 
$\sim 10^{-4}M_{\odot}/{\rm yr}$, while $c_{\rm s}^3/G$ becomes 
large owing to the higher temperature in the outer part.
Note that such an outer portion is outside the central cooled region 
since the dark-matter gravity is dominant.  

Recently, O'Shea et al. (2005) studied the evolution of a fossil HII
region from a similar initial condition to ours.
Their conclusion agrees with ours in the respect that
the enhanced electron fraction promotes H$_2$ formation
and the resultant lower temperature enables formation of lower mass stars 
than the first-generation stars.
Since they utilized detailed three-dimensional hydrodynamics, 
their calculation is superior to ours with respect to the dynamics.
However, they did not include deuterium chemistry, which is found to be
important in our calculation.
In this regard, our calculation is complementary to theirs.
They discussed the mass scale of the second-generation stars
by comparing the accretion timescale $t_{\rm acc}=M_{\ast}/\dot{M}$ with
the Kelvin-Helmholtz timescale $t_{\rm KH}=GM_{\ast}^2/R_{\ast}L_{\ast}$: 
since it takes a Kelvin-Helmholtz timescale for a protostellar
cloud to contract enough to ignite nuclear fusion, 
the mass where these two timescales become equal corresponds to the 
mass of an accreting protostar when it reaches the main sequence.
Note that in deriving the mass-scale of the stars by equating these 
two timescales, O'Shea et al. (2005) implicitly assumed that 
the accretion is halted by an unknown mechanism at the onset 
of the hydrogen burning.
For comparison with O'Shea et al. (2005), 
if we adopt the same timescale argument, the stellar mass becomes 
$\sim$ 16 M$_{\odot}$ in our case, with the timescales 
$\sim$ 2.2 $\times$ 10$^5$ yr.
In this evaluation, we used the stellar parameters listed in Schaerer (2002).
This stellar mass-scale is similar to that by O'Shea et al. (2005) 
owing to the similar accretion rate to theirs ($\sim 10^{-4}
M_{\odot} {\rm yr}^{-1}$ ).
This lower accretion rate compared with that of the first stars 
($10^{-2}-10^{-3}M_{\odot} {\rm yr}^{-1}$) is due to lower temperature 
by the HD cooling in our case, while in O'Shea et al. (2005) 
it is due to high angular momentum of the star forming clumps.
If we included both of these effects, the accretion rate and then 
the stellar mass-scale would be even lower.

Johnson \& Bromm (2005) and Shchekinov \& Vasiliev (2005) 
stressed the importance of HD chemistry in the early universe,
studying evolution of gas ionized by strong shocks.
This situation is similar to that supposed by Uehara \& Inutsuka (2000). 
Although the cause of ionization is different from ours, the final
consequence is similar: the HD formation is facilitated by ionization 
and the temperature declines as low as the CMB temperature.
Johnson \& Bromm (2005) also mentioned the evolution of 
fossil HII regions and pointed out the importance of HD.
They suggested that the second-generation stars have lower mass
than the first-stars owing to the lower temperature environment. 
We here confirmed their claim by more detailed calculations.

In this paper, we do not include external radiative heating, except 
for the CMB radiation in our calculation.
By comparing runs with and without CMB radiation, we find that
the sole effect of the CMB radiation is preventing the gas from falling 
below the radiation temperature: it does not alter the higher temperature 
behavior at all.
If the first stars have already created the UV background radiation, we need
to take account of such effects as photoionization/dissociation and 
photo-heating.
Oh \& Haiman (2003) discussed that in the presence of a UV radiation field,
subsequent star formation is prevented in fossil HII regions.
As a future work, we are planning to include effects of the UV field in our
calculation and examine the validity of their claim.

Finally,
we need to consider the effect of a remnant BH.
The radiation from the gas accreted onto a BH can affect the surrounding gas.
However, O'Shea et al. (2005) indicated that photodissociation flux
expected from this 
is insufficient to alter the H$_2$ fraction significantly.   
The BH also has an effect on the gas dynamics through the gravitational
field.
The length scale of the region influenced by the BH gravity is given by the
Bondi accretion radius 
\begin{equation}
r_{{\rm Bondi}} \sim
3.3 \times 10^{-3} \left( \frac{T}{10^4 {\rm K}} \right)^{-1}
\left( \frac{M_{{\rm BH}}}{100 {\rm M}_{\odot}} \right)
\left( \frac{\mu}{0.6} \right) \quad {\rm pc} ,
\end{equation}
where M$_{{\rm BH}}$ and $\mu$ are the BH mass and the mean molecular
weight of the gas, respectively.
The mass inside the Bondi radius is initially negligible 
compared to the total mass of the collapsing gas.
With cooling and contraction, however, the Bondi radius becomes as large 
as the central core radius.
Therefore if the BH exists exactly at the center of the halo, 
all the cooled gas is eventually accreted into the BH, 
instead of forming the next generation of stars.
However, even in the case of direct BH formation, part of the mass is 
likely to be shed non-spherically.
If so, the BH would be kicked out of the shallow potential wells of
small halos, or even if it could remain inside the halo, it would be
off-center.
In these cases, the second-generation star formation occurs as shown in 
our calculation.

\section{Summary}

We have investigated the evolution of photo-ionized gas in small halos
since the death of the exciting first star, by using one-dimensional 
hydrodynamics incorporating primordial gas chemistry.
We have found that the high electron fraction induces molecular hydrogen
formation in a larger amount than in a cloud without initial ionization.
The enhanced H$_2$ cooling makes temperature in the central region lower
than the threshold value for HD formation.
The cooling by HD further drops the temperature below 100 K.
We have also shown that the gas in halos with $\ga$ 10$^6$
M$_{\odot}$ cools sufficiently and subsequently forms stars.
In those HD-cooling halos, the gas collapses further at $<$ 100 K and
eventually fragments into low-mass cores. 
Inside these cores, low-mass stars are expected to form.
The nature of second-generation stars can be quite different from that 
of first stars despite being composed of the same material.

\section*{Acknowledgments}
The authors would like to thank Tom Abel for fruitful discussion, 
Toshitaka Kajino for continuous encouragement, and 
the referee, Emanuele Ripamonti,  
for valuable comments improving the manuscript.
This work is supported in part by the Grants-in-Aid
by the Ministry of Education, Science and Culture of Japan
(16204012:KO).

\end{document}